\documentclass[showpacs,preprintnumbers,amsmath,amssymb]{revtex4}
\usepackage{tabularx,graphicx}

\usepackage{color}
\usepackage{hyperref}
\hypersetup{
    colorlinks=true,
    linkcolor=blue,
    filecolor=blue,      
    urlcolor=blue,
}

\begin{document}

\newcommand{\beq}{\begin{equation}}
\newcommand{\eeq}{\end{equation}}
\newcommand{\beqn}{\begin{eqnarray}}
\newcommand{\eeqn}{\end{eqnarray}}
\newcommand{\bmath}{\begin{subequations}}
\newcommand{\emath}{\end{subequations}}
\newcommand{\bra}[1]{\langle #1|}
\newcommand{\ket}[1]{|#1\rangle}

\title{Entropy generation  and momentum transfer in the superconductor-normal  and normal-superconductor
phase transformations
and the consistency of the conventional theory of superconductivity}
\author{J. E. Hirsch }
\address{Department of Physics, University of California, San Diego,
La Jolla, CA 92093-0319}

\begin{abstract} 
Since the discovery of the Meissner effect the superconductor to normal (S-N)  phase transition  in
the presence of a magnetic field  is understood to be a first order phase transformation that is reversible under ideal conditions and
obeys the laws of thermodynamics. The reverse (N-S) transition is  the Meissner effect. This implies in particular that the kinetic energy of the supercurrent is not dissipated
as Joule heat in the process where the superconductor becomes normal and the supercurrent stops.
In this paper we analyze the entropy generation and the momentum transfer between the supercurrent and the body
 in the S-N transition  and the N-S transition 
as described by the conventional theory of superconductivity. We find that it is not possible to explain 
the transition in a way that is consistent with the laws of thermodynamics unless the momentum transfer between the
supercurrent and the body occurs with zero entropy generation, for which the conventional theory of
superconductivity provides no mechanism. Instead, we point out   that  the alternative theory of hole superconductivity 
does not encounter such difficulties.
 \end{abstract}
\pacs{}
\maketitle 

\section{introduction}
The conventional theory of superconductivity \cite{londonbook,schriefferbook,tinkham} is generally believed to describe both the Meissner
effect and its reverse, the superconductor-normal (S-N) transition in the presence of a magnetic field,
in a way that is consistent with the laws of  thermodynamics that describe reversible first order
phase transformations. That the transition is reversible has been established long ago. It was strongly suggested
by the discovery of the Meissner effect \cite{meissner}, was first proposed theoretically by Rutgers \cite{rutgers} and by 
Gorter and Casimir \cite{gortercasimir}, and was confirmed experimentally by extensive experimental work  by Keesom and
coworkers \cite{keesom} and Mapother \cite{mapother}. It is also an implicit assumption of the conventional theory of 
superconductivity \cite{londonbook,schriefferbook,tinkham}. A detailed discussion of this issue is given in the book by Shoenberg  \cite{shoenberg}.

We have recently pointed out  \cite{revers,momentum} that there are key issues regarding basic conservation laws raised by the fact that
the transition is reversible that   have not
been discussed in the theoretical  superconductivity literature. We have discussed how these
issues are addressed and resolved in an alternative theory of superconductivity, the theory of
hole superconductivity \cite{holesc},  and we have raised the possibility that 
the conventional theory of superconductivity may lack essential physical ingredients that are necessary to describe
these transitions in a way that is consistent with the laws of thermodynamics, electrodynamics and mechanics
 \cite{lorentz,lenz,missing,meissnerorigin,revers,momentum}.
 
 This paper deals with type I superconductors only. A superconductor in a magnetic field has a surface supercurrent that prevents the magnetic field from penetrating its interior. The supercurrent carries mechanical 
 momentum and kinetic energy. When the system undergoes a transition to the normal state, the supercurrent stops since the normal state does not carry any current. In the reverse transition, when the normal metal in a magnetic field becomes
 superconducting, a supercurrent carrying mechanical momentum and kinetic energy spontaneously starts flowing. 
In this paper we explore issues associated with entropy generation and momentum transfer associated with these transformations and ask  whether
the conventional theory of superconductivity has the ability to account for these processes in a consistent way. 
We compare the superconductor-normal transition to the liquid-vapor transition to clarify some of the issues at play. 
We find that the conventional theory of superconductivity cannot describe the S-N  transition nor the N-S transition in the presence of a magnetic field
in a consistent way. In particular, we find that  it is required that {\it the momentum of the supercurrent is transferred to the body
as a whole without any entropy generation}, a process for which the conventional theory provides no mechanism.  
We  find that in the conventional description the transition generates entropy even when it proceeds infinitely slowly, in contradiction with the
laws of thermodynamics. We also find that it violates energy conservation.
We furthermore find that thermodynamic equilibrium between the normal and superconducting phases is not  properly
described. We also find that the conventional theory predicts that as the temperature goes to zero the 
time it takes for the transition to take place diverges, which is inconsistent with observations. We conclude that the conventional theory cannot describe
the transition in a consistent way. We briefly sketch the way the unconventional theory of hole superconductivity describes the transition in a way
that is consistent with physical laws and with experiment.

In more detail, the outline of this paper is at follows. In Sects. II and III, we discuss the general question of entropy production in a  first order 
phase  transition,
using the liquid-gas transition as a well-understood model. In Sect. IV we apply the same reasoning to the 
superconductor-normal transition in a magnetic field, and show that the entropy production predicted by thermodynamics
is entirely accounted for by the Joule heat generated by eddy currents in the transition. The remainder of the paper  analyzes
whether or not the momentum transfer between the supercurrent and the body, that has to take place to 
respect momentum conservation, can be described within the conventional theory  without
additional entropy production. Sect. V quantifies the momentum that needs to be transferred, showing that it is
several orders of magnitude larger than would be expected at first sight, as well as its associated  kinetic energy, showing
that it is a bulk rather than a surface effect. Stringent bounds set by experiment determine that essentially none
of this kinetic energy is dissipated as Joule heat. The following sections (VI - XV) explore whether within the
physics of the transition as described by the conventional theory of superconductivity it is possible to account for 
the momentum transfer without introducing irreversibility, and conserving momentum and energy. We find that it
is not possible. A key part of the argument is to show that the normal electron distribution that results from pair dissociation is necessarily anisotropic, 
as shown in Figs. 6  and 11. In Sect. XVI we summarize our findings and discuss their implications. Sect. XVII briefly
explains how the theory of hole superconductivity can explains these questions, and we conclude with closing
arguments in Sect. XVIII.

The questions discussed in this paper have never been discussed in the superconductivity literature before. 
Earlier theoretical work on the kinetics of the normal-superconductor transition in a magnetic field \cite{dorsey,goldenfeld} 
used the
 time-dependent Ginzburg-Landau (TDGL) formalism \cite{tdgl1,tdgl2,cyrot}.
That formalism is phenomenological and  
involves a {\it first order} differential equation in time with $real$ coefficients for the time evolution of the order parameter. Hence it describes
$irreversible$ time evolution, and is therefore irrelevant to the Meissner effect for type I superconductors,
which is a $reversible$ process \cite{shoenberg}, as discussed above. 
There is also theoretical work 
in the literature on the
resistive transition in a magnetic field \cite{resist} describing the onset of resistance in type II superconductors
 through the creation of phase slips at a finite rate. 
Such treatments are also  not relevant to the physical situation considered in this paper. For our case, a simply  connected type I superconductor
in the presence of a magnetic field, the phase of the order parameter is uniform in the superconducting phase
and the transition to the normal state does not occur through phase slips or phase fluctuations
 but rather through suppression of the amplitude of the
superconducting order parameter.

Superconductivity is undoubtedly a quantum phenomenon. BCS theory undoubtedly describes many of the quantum aspects of superconductivity correctly, for example
the existence of macroscopic phase coherence and associated Josephson phenomena. The treatment in this paper is certainly not a fully quantum treatment, 
it may be characterized as semiclassical. However, whether quantum, semiclassical or classical, a physical description of nature has to satisfy fundamental
laws of physics such as momentum and energy conservation. The author can see no way in which a fully quantum description of the phenomena discussed here 
within the conventional theory would be able to
resolve the questions raised here, without introducing new physics such as electron-hole asymmetry, which is not contained in BCS theory and is
contained in the theory of hole superconductivity. 

\section{thermodynamics of first order phase transformations}

Consider the generic situation shown in Figure 1. The ``universe'' that can potentially change its entropy consists of our
system (small box) initially at temperature $T$ and a large heat bath at temperature $T+\Delta T$. The heat bath is sufficiently large that its temperature 
doesn't change when it gives or takes  heat to or from the system.
The system is initially   in its lower entropy phase (phase 1).  We assume its initial temperature T is the temperature
of coexistence of phase 1 with another phase (phase 2) of higher entropy for a given other parameter X, i.e. T=T(X). Specifically,
for the liquid-gas transition X is the pressure P, for the S-N transition X is the applied magnetic field H.

We will ignore questions related to the initial stages of the transition involving nucleation, surface energy of domains, and superheating. In other words, we assume in the initial state when the system is in phase 1 there is already a coexisting small amount of phase 2 that can grow without having to overcome a barrier.

When thermal contact is established, heat will flow from the heat bath to the system. At the end of the process, the system
will be in phase 2 at temperature $T+\Delta T$. Let us call $Q$ the net amount of heat that flowed from the heat bath to the system in this process.
The change in entropy of the universe is then
\beq
\Delta S_{univ}=-\frac{Q}{T+\Delta T}+\frac{Q}{T}=\frac{Q}{T}\frac{\Delta T}{T}
\eeq
to lowest order in $\Delta T$.

          \begin{figure}
 \resizebox{8.5cm}{!}{\includegraphics[width=6cm]{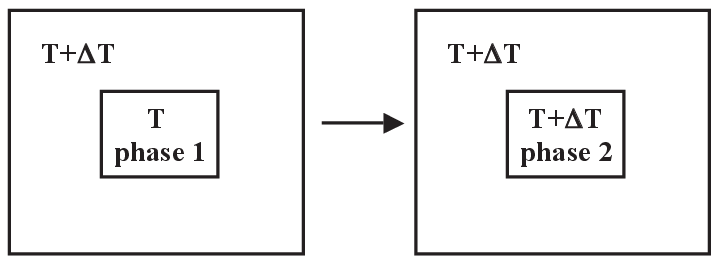}}
 \caption { Schematics of a first order phase transformation of a system (small box) in thermal contact with 
 a heat bath (large box) making a transition from a lower entropy phase (phase 1) to a higher entropy phase (phase 2). 
 We assume an external parameter (pressure or magnetic field) has the value corresponding to phase equilibrium for the
 system at temperature $T$.
 }
 \label{figure1}
 \end{figure}  

Let us call $L(T)$ the latent heat of the phase transformation for the given amount
of system at temperature $T$,  and $C_1$, $C_2$ the heat capacities of phases 1, 2. The latent heat is determined by the difference in entropy of the two
coexisting phases
\beq
L(T)=T(S_2(T)-S_1(T)) .
\eeq
Assume the system absorbs the latent heat while it is at temperature $T$, transforming
to phase 2, and then raises its temperature to $T+\Delta T$.
The heat transferred from the bath to the system is then
\beq
Q=L(T)+C_2\Delta T
\eeq
and the change in entropy of the universe to first order in $\Delta T$ is from Eq. (1),
\beq
 \Delta S_{univ}= \frac{L(T)}{T}\frac{\Delta T}{T}
 \eeq
 Alternatively, we could assume that the system first raises its temperature to $T+\Delta T$ while remaining in phase 1 and then undergoes the
 transformation to phase 2. The calculation would be slightly different but the end result for the change in entropy of the universe still has to be equation (4),
 because entropy is a function of state. We will discuss specific examples   in the next sections.

\section{the liquid-gas phase tranformation}

     \begin{figure}
 \resizebox{6.5cm}{!}{\includegraphics[width=6cm]{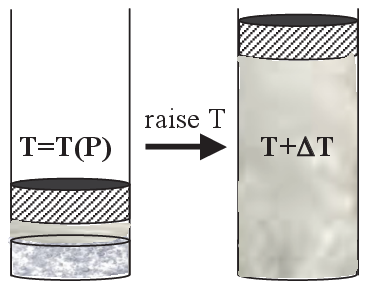}}
 \caption { Liquid in equilibrium with vapor at temperature $T(P)$ and pressure $P$ exerted by a piston (dashed disk) of weight
 $P\times A$, with $A$ the cross-sectional  area. The temperature is raised to $T+\Delta T$, the liquid turns into vapor and the piston rises.
 }
 \label{figure1}
 \end{figure} 
Our system, shown in Fig. 2,  consists of 1 mole of a liquid (e.g. water) in a cylinder with a piston of mass M and area A exerting pressure P=Mg/A. Assume $P=P(T)$ is the pressure of phase equilibrium between liquid and gas at temperature $T$, which
satisfies the Clausius-Clapeyron equation  
\beq
\frac{dP}{dT}=\frac{L(T)}{T\Delta V}
\eeq
with $\Delta V$ the difference in molar volumes of liquid and gas. In the phase transformation at
temperature $T$ the system absorbs heat $L(T)$ and performs work $P\Delta V$.

Let us assume there is a small  friction force between the piston and the cylinder wall, and we ignore any difference between static and kinetic friction.
Because of the friction force  the piston
will only move if there is additional pressure $\Delta P$ in the liquid, and we assume that the point
($P+\Delta P$, $T+\Delta T$) is still on the coexistence line. The system will first absorb
heat $Q_1$ from the bath while still liquid,
\beq
Q_1=C_1\Delta T
\eeq
with  change in entropy of the universe $O(\Delta T^2)$.  Then, the system undergoes the phase transformation at temperature $T+\Delta T$ against
pressure $P+\Delta P$, absorbing heat $L(T+\Delta T)$ from the heat bath. The heat transfer generates no entropy since it happens
between the bath and the system  at the same temperature.
 In the process of converting to gas the system does work $P\Delta V$ and dissipates friction heat
\beq
Q_{frict}=\Delta P \Delta V  .
\eeq
We can think of this friction heat equivalently as going back to the heat bath, increasing the bath's entropy,
or as being used as part of the $L(T+\Delta T)$,  so that the heat bath supplies less heat hence decreases
its entropy by less. Either way, the change in entropy of the universe to order $\Delta T$  is
\beq
\Delta S_{univ}=\frac{Q_{frict}}{T}=\frac{L(T)}{T}\frac{\Delta T}{T}
\eeq
where we have used the Clausius-Clapeyron eq. (5). Eq. (8) is in agreement with Eq. (4).

The net heat transferred from the bath to the system in this process was
\beq
Q=C_1\Delta T+L(T+\Delta T)-\Delta P \Delta V
\eeq
which is the same as Eq. (3), since from Eq. (5), $\Delta P \Delta V=(L(T)/T)\Delta T$, 
and from expanding $L(T+\Delta T)$ and using Eq. (2) for $dL/dT$,
\beq
L(T+\Delta T)=L(T)+L(T)\frac{\Delta T}{T}+T(C_2-C_1)\Delta T 
\eeq
to first order in $\Delta T$.

Alternatively, if there is negligible friction between the piston and the cylinder, the piston will accelerate upward and acquire kinetic energy
\beq
K_{piston}=\Delta P \Delta V
\eeq
when all the liquid has evaporated, and will oscillate up and down, eventually dissipating this energy as heat which goes back to the
heat bath, giving a change in the entropy of the bath (and the universe)
\beq
\Delta S=\frac{K_{piston}}{T}=\frac{L(T)}{T}\frac{\Delta T}{T}
\eeq
in agreement with Eq. (4).

In this latter process assuming negligible friction between the piston and the cylinder, the increase in the entropy of the universe
will approach zero as the temperature difference $\Delta T$ goes to zero, in which case the transition will
proceed increasingly slower: the speed of the piston goes as $v\sim \sqrt{(2/M)K_{piston}}$, and as
$\Delta T\rightarrow 0$, $\Delta P$ and $K_{piston}\rightarrow 0$. The converse however is not true: the transition may proceed
infinitely slowly and yet give rise to a finite increase in the entropy of the universe, as would be the case
for finite friction force between the piston of the cylinder $F_{friction}$ for the case where  $\Delta P$ is infinitesimally larger  than 
$F_{friction}/A$.

     \begin{figure}
 \resizebox{6.5cm}{!}{\includegraphics[width=6cm]{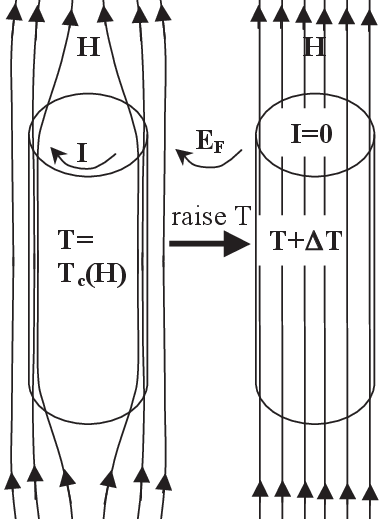}}
 \caption { Superconducting cylinder in magnetic field $H=H_c(T)$ becoming normal when $T$ is raised to $T+\Delta T$.
 }
 \label{figure1}
 \end{figure}

\section{the  superconductor-normal phase transformation}
Consider a superconducting cylinder at temperature $T$  in a magnetic field $H=H_c(T)$, where $H_c(T)$ is the
thermodynamic critical field at temperature $T$. When the temperature is slightly raised to
$T+\Delta T$ the system will become normal. In many ways this transition, shown in Figs. 3 and 4, is similar to the liquid-vapor transition
shown in Fig. 2. In particular, the same thermodynamic laws should apply. The analogous of the Clausius-Clapeyron equation 
(5) is \cite{reif}
\beq
\frac{dH_c}{dT}=\frac{L(T)}{T(M_s-M_n)}
\eeq
where $M_n=0$ and $M_s=-H_c /4\pi$ are the magnetization densities  in the normal and superconducting states.
Hence from Eq. (13)
\beq
\frac{L(T)}{T}=-\frac{H_c}{4\pi}\frac{dH_c}{dT}
\eeq
and the change in entropy of the universe predicted by thermodynamics when the superconducting cylinder goes normal is, from Eqs. (4) and (14)
\beq
\Delta S_{univ}=-\frac{H_c}{4\pi} \frac{dH_c}{dT}\frac{\Delta T}{T}  .
\eeq

      \begin{figure}
 \resizebox{6.5cm}{!}{\includegraphics[width=6cm]{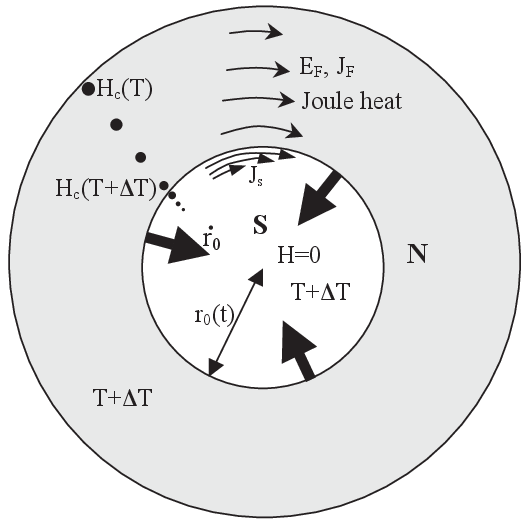}}
 \caption {Superconducting cylinder becoming normal, viewed from the top. The phase boundary moves in with velocity $\dot{r}_0(t)$.
 The changing magnetic flux generates a Faraday field $E_F$, giving rise to eddy current density $J_F=\sigma E_F$, with $\sigma$ the
 normal state conductivity, and Joule heat is dissipated.
 }
 \label{figure1}
 \end{figure} 
 
 Note that the superconductor-normal transition in a magnetic field cannot occur through nucleation of ``bubbles'' of the normal phase in the interior of the superconducting region, as other first-order phase transformations occur. This is because in the interior 
 there is no magnetic field and we are assuming that the temperature is well below $T_c$. The transition has to occur
 through the normal phase moving inward from the surface, as the region carrying the Meissner current becomes normal.
 This is clearly explained in the theoretical and experimental works of London \cite{londonh}, Pippard \cite{pippard}
  and Faber \cite{faber}. In particular, Faber \cite{faber} argues theoretically and shows experimentally that in
  a cylindrical geometry the superconducting phase can collapse inwardly retaining near cylindrical symmetry at all times.
  For simplicity we will assume perfect cylindrical symmetry in   this paper except when explicitly stated
  otherwise (Fig. 10).
 
What is the physical origin of the entropy increase Eq. (15)? When the system goes normal the magnetic field enters the body, and the
changing magnetic flux generates eddy currents in the normal region that decay by dissipation of Joule heat. 
That is the analogous of the piston friction discussed in the previous section. Here we will show that this physics accounts for the
entire entropy increase predicted by thermodynamics Eq. (15).
Let us define
\bmath
\beq 
H_c \equiv H_c(T+\Delta T)
\eeq
\beq
H_c(1+p) \equiv H_c(T)
\eeq
hence
\beq
p=-\frac{1}{H_c(T)}\frac{dH_c(T)}{dT}\Delta T  
\eeq
\emath
and substituting in Eq. (15)
\beq
\Delta S_{univ}=\frac{H_c^2}{4\pi T} p   .
\eeq

Consider the conservation of energy equation that follows from Faraday's law $\vec{\nabla} \times \vec{E}=(-1/c)\partial \vec{H} /\partial t$ and
Ampere's law $\vec{\nabla} \times \vec{H}=(4\pi /c) \vec{J}$:
\beq
\frac{d}{dt}(\frac{H^2}{8\pi}) = -\vec{J}\cdot \vec{E} - \frac{c}{4\pi} \vec{\nabla}\cdot (\vec{E}\times \vec{H}).
\eeq
 The left side represents the change in energy of the electromagnetic field as the magnetic field $H_c(T)$ enters the body, the first term on the right side is the work done by the electromagnetic
 field in creating currents in this process, and the second term is the inflow of electromagnetic energy. Integrating over the volume of the body $V$  and over time we find for the
 change in electromagnetic energy per unit volume
 \beq
\frac{1}{V} \int d^3r \int_0^\infty dt \frac{d}{dt}(\frac{H^2}{8\pi}) =\frac{H_c^2 (1+p)^2}{8 \pi} .
 \eeq
 since initially the magnetic field is completely excluded from the body.
 From Faraday's law and assuming cylindrical symmetry we have for the electric field generated by the changing magnetic flux at the surface
 of the cylinder
 \beq
 \vec{E}(R,t)=-\frac{1}{2\pi R c}\frac{d}{dt}  \phi(t) \hat{\theta}
 \eeq
 where $R$ is the radius of the cylinder and $\phi(t)$ is the magnetic flux throught the cylinder, with
 $\phi(t=0)=0$, $\phi(t=\infty)=\pi R^2 H_c (1+p)$. 
Integration of the second term on the right in Eq. (18), the energy inflow, over space and  time,
converting the volume integral to an integral over the surface   of the cylinder, 
using that $H=H_c(1+p)$ at the surface of the cylinder independent of time and Eq. (20)
for the electric field at the surface yields
 \beq
 \frac{1}{V} \int_0^\infty dt    \oint (- \frac{c}{4\pi}) (\vec{E}\times \vec{H}) \cdot d\vec{S}=\frac{H_c^2 (1+p)^2}{4 \pi} .
 \eeq
 for the total electromagnetic energy flowing in through the surface of the sample during the
 transition.
 
 The current $\vec{J}$  in Eq. (18) flows in the azimuthal direction and is given by the sum of superconducting and normal currents
 \beq
J(r)=J_s(r)+J_n(r)
\eeq
where $J_s(r)$ flows in the region $r\leq r_0(t)$ and is of appreciable magnitude only within $\lambda_L$ of the phase boundary, where $\lambda_L$
is the London penetration depth. $r_0(t)$ is the radius of the phase boundary (see Fig. 4) at time $t$. 
Integration of the second term in Eq. (18) over the superconducting current yields \cite{pp2,londonh}
\beq
\frac{1}{V} \int d^3r \int_0^\infty dt  (-\vec{J}_s \cdot \vec{E} )=-  \frac{H_c^2}{8\pi}  .
\eeq
This is because the Faraday field accelerates the supercurrent \cite{pp2} until its kinetic energy density becomes the difference in free energies
between the normal and the superconducting states, at which point the supercurrent stops \cite{londonh}.

The Joule heat per unit volume generated during the transition is
\beq
Q_J \equiv \frac{1}{V} \int d^3r \int_0^\infty dt  \vec{J}_n \cdot \vec{E}
\eeq
hence from integrating Eq. (18) over space and time using Eqs.  (19), (21), (22) and (23) we have
\beq
\frac{H_c^2 (1+p)^2}{8\pi}=-  \frac{H_c^2}{8\pi} - Q_J + \frac{H_c^2 (1+p)^2}{4\pi}
\eeq
which implies 
\beq
Q_J = \frac{H_c^2}{4\pi}p 
\eeq
to linear order in $p$. The entropy generated from Joule heat is then
\beq
\Delta S_{Joule}=\frac{Q_J}{T}= \frac{H_c^2}{4\pi T} p
\eeq
and comparing Eqs. (26) and (17)
\beq
 \Delta S_{univ}=\Delta S_{Joule}  .
 \eeq
 We conclude that the change in entropy of the universe in the superconductor-normal
 phase transformation results $solely$ from the Joule heat $Q_J$ Eq. (24) produced by normal 
 current $J_n$  generated during the transition.
 
 So far we have not made any assumption about the physical origin of the normal
 current $J_n$. The Faraday electric field created by the changing magnetic flux
 will generate a Faraday current
 \beq
 \vec{J}_F(r,t)=\sigma \vec{E}_F(r,t)
 \eeq
 with 
 \beq
 \sigma=\frac{n e^2 \tau}{m_e}
 \eeq
 the normal state conductivity, with $n$ the normal state carrier density. Following the calculation in refs. \cite{pippard,pp2} the
 Faraday electric field is given by
   \beq
 \vec{E}_F(r) = \frac{r_0}{cr} \dot{r}_0H_c \hat{\theta}
 \eeq
 to lowest order in $p$,
 pointing in the azimuthal direction $\theta$ (counterclockwise). $\dot{r}_0(t)$ is the velocity of motion of the phase boundary.
The energy per unit volume dissipated in the transition due to the
Faraday current Eq. (29)  is given by
 \beq
 W=\frac{1}{\pi R^2}\int_0^{t_0} dt\int_{r_0(t)}^0dr (2 \pi r) J_F(r,t) E_F(r,t) .
 \eeq
 $t_0$ is the total time for the transition, given by  (for small p)  \cite{pippard,pp2}
 \beq
 t_0=\frac{\pi \sigma}{p c^2} R^2 .
 \eeq
Eq. (32) yields using Eqs.  (29) and  (31) 
\beq
W=\frac{2\sigma H_c^2}{R^2 c^2}\int_0^{t_0} dt r_0^2 ln(\frac{R}{r_0})   \dot{r}_0^2
=\frac{2\sigma H_c^2}{R^2 c^2}\int_R^0dr_0 r_0^2  ln(\frac{R}{r_0}) \dot{r}_0
\eeq 
and using that \cite{pippard,pp2}
\beq
r_0 \dot{r}_0ln(\frac{R}{r_0})=-\frac{pc^2}{4\pi \sigma}
\eeq
yields
\beq
W=\frac{H_c^2}{4\pi}p=Q_J .
\eeq
Therefore, we conclude that the Joule heat Eq. (24) is produced by the
Faraday current Eq. (29).

 From Eq. (33) we learn that   the increase in the entropy of the universe is inversely proportional
to the time for the transition to take place. In terms of the average velocity of motion of 
the phase boundary $v=R/t_0$ the increase in entropy Eq. (17) can be written, using  Eq. (33), as
\beq
\Delta S_{Joule}=\frac{H_c^2}{16\pi T} (\frac{R}{\lambda_L})(\frac{\ell}{\lambda_L})
(\frac{n}{n_s} )\frac{v}{v_F}
\eeq
with $v_F$ the Fermi velocity, $\sigma$ given by Eq. (30), $\ell=v_F\tau$ the mean free path,
$n_s$ the superfluid density,  and
the London penetration depth $\lambda_L$  given by the usual expression \cite{tinkham}
\beq
\frac{1}{\lambda_L^2}=\frac{4\pi n_se^2}{m_e c^2} .
\eeq

 In summary, we find that the increase in entropy predicted by thermodynamics for the S-N transition  is completely accounted for by the
 entropy generated by the Joule heat resulting from the decay of the eddy currents driven by
 the Faraday electric field. Exactly the same analysis holds for the reverse transition from normal to superconducting 
 state (Meissner effect). 
  In the following sections we analyze the processes of
 momentum transfer from the supercurrent to the body and vice versa   and whether or not they generate additional entropy.
 
 \section{momentum of the supercurrent}
 
 Consider the process where the superconductor goes normal. Initially a supercurrent circulates within a London penetration depth
of the surface, $\lambda_L$.
The supercurrent  density  
 when the magnetic field is $H_c$ is                                                      
 \bmath   
 \beq
 J_s=n_s e v_s=\frac{n_s e^2\lambda_L} {m_ec}H_c
 \eeq
 with $v_s$ the superfluid velocity. 
 Associated with it is a mechanical momentum density \cite{rostoker}
 \beq
 \mathcal{P}_s=-\frac{m_e}{e}J_s
  \eeq
  \emath
  with $m_e$ the bare electron mass. The total angular momentum of the supercurrent is
  \beq
  L_e= \mathcal{P}_s R(2\pi R \lambda_L h)=-\frac{m_e c}{2e}R^2hH_c
  \eeq
  where we have used Eq. (38) in the second equality.
  This mechanical momentum needs to be transferred to the body as a whole
  in a reversible way when the supercurrent stops.

Alternatively we can derive Eq. (40) as follows. The magnetization that nullifies the magnetic field in the
interior is given by
\beq
\vec{M}=-\frac{\vec{H_c}}{4\pi} .
\eeq
The associated magnetic moment is
\beq
\vec{\mu}=\vec{M}\pi R^2 h=-\frac{R^2h}{4}\vec{H_c} .
\eeq
Using the relation between magnetic moment and orbital angular momentum $\vec{L}_e$ of the electrons in the supercurrent 
\beq
\vec{\mu}=\frac{e}{2m_ec}\vec{L}_e
\eeq
we obtain
\beq
\vec{L}_e=-\frac{m_e c}{2e}R^2h\vec{H_c} 
\eeq
as in Eq. (40).

The kinetic energy associated with the initial supercurrent is
\beq
K_e(R)\equiv K_e=\frac{m_e J_s^2}{n_s e^2}(2\pi R\lambda_L h) =\frac{H_c^2}{8\pi}(2\pi R\lambda_L h)
\eeq
where $h$ is the height of the cylinder.
Note that the angular momentum of the supercurrent Eq. (44) is proportional to the volume of the system while its
kinetic energy Eq. (45)  is proportional to the surface area.

The transfer of the electronic angular momentum  to the body also entails  transferring some kinetic energy to the body, however the kinetic energy of the body associated
with angular momentum Eq. (44) is only a tiny fraction of Eq. (45) 
because of the large body mass (of order $(m_e/m_{ion})(\lambda_L/R)$) and hence can be disregarded.

It may appear that the angular momentum Eq. (44) is all the angular momentum that needs to be transferred to the body
when the system becomes normal.
Then, it could be argued \cite{referee} that when the system becomes  normal and finite resistivity sets in, the supercurrent 
that flows within $\lambda_L$ of the surface of the body Eq. (39a) will decay by collisions with impurities and/or phonons,  its angular momentum Eq. (44) would be transmitted to the body by those collisions
making the body rotate, and its kinetic energy Eq. (45) would be dissipated
as Joule heat. If so, the Joule heat dissipated per unit volume would be
\beq
Q=\frac{K_e}{\pi R^2 h}=\frac{H_c^2}{8\pi}\frac{2\lambda_L}{R}
\eeq
which is a small fraction of the Joule heat dissipated by eddy currents Eq. (26), negligible for macroscopic samples.

            \begin{figure}
 \resizebox{8.5cm}{!}{\includegraphics[width=6cm]{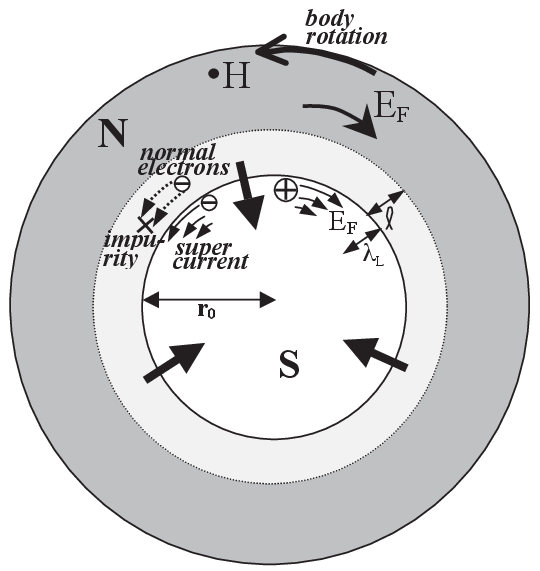}}
 \caption {Superconducting cylinder becoming normal, as seen from the top. Magnetic field points out of the page.
 According to the conventional theory, as electrons in the supercurrent enter the normal region they scatter off impurities and transfer their momentum
 to the body as a whole, generating the body rotation in direction opposite to the force exerted by the Faraday field
 $E_F$ on the positive ions. }
 \label{figure1}
 \end{figure} 
 
However,  this is not so.
The reason is that the transition cannot occur by the supercurrent near the surface simply stopping. Instead, the phase
boundary moves inward as shown in Figs. 4 and 5. 
For simplicity we assume that the phase boundary has cylindrical symmetry at all times.
As the supercurrent near the surface stops, supercurrent
further inward is generated. As the magnetic field and the supercurrent move inward,
 the Faraday field transfers angular momentum to the body in clockwise direction, opposite to
the angular momentum Eq. (44) that the body has to acquire. The total angular momentum transferred by
the Faraday field to the body in the transition  is
\beq
\vec{L}_F=e\int_0^\infty dt n_s(2\pi r_0\lambda_L h)\vec{r}_0\times \vec{E}_F(r_0)=-\frac{R}{3\lambda_L}\vec{L}_e ,
\eeq
which is $much$ larger than $L_e$. 
So e.g. for $\lambda_L=400\AA$, $R=1.2cm$,  angular momentum $(10^5+1)\vec{L}_e$ has to be transferred to the body
through another mechanism for the body to end up with the correct angular momentum $\vec{L}_b=\vec{L}_e$.

The total kinetic energy associated with the supercurrents generated during this process is
\beq
K_e^{tot}= \int_0^R \frac{dr_0}{\lambda_L} K_e(r_0) =   \frac{H_c^2}{8\pi} (\pi R^2h) 
\eeq
which  is the difference in free energies between the normal and superconducting states.  This energy cannot be dissipated as Joule heat since the transition is reversible. 
The latent heat $L$ associated with the superconductor-normal transition at temperature $T$ is
\beq
L=4\frac{(T/T_c)^2}{1-(T/T_c)^2}K_e^{tot} .
\eeq
Experimentally it is found that less than $1\%$ of the latent heat Eq. (49) is dissipated as Joule heat
in the transition \cite{keesom,mapother}. The thermodynamic relations are found to hold even down to
 temperatures below $0.1T_c$ \cite{finn}, where   the total  kinetic energy of the supercurrent  Eq. (48)  is more than 20 times the latent heat. This confirms that the kinetic energy of the supercurrent is not dissipated when the
supercurrent stops, rather it is stored in the electronic degrees of freedom, to be retrieved again in the
reverse transition when the system goes superconducting and expels the magnetic field.
The question then is: how is the electronic angular momentum Eq. (47), that has associated with it the kinetic
energy Eq. (48), transferred to the body without any dissipation?
 
 \section{the transition according to the conventional theory}

  Figure 5 shows schematically how the transition  is envisioned to happen within the conventional theory of superconductivity \cite{eilen,halperin}. As the N-S phase boundary moves inward, Cooper pairs at the edge of the S-region dissociate and the resulting normal electrons
  inherit the center of mass momentum of the Cooper pair, without inheriting the kinetic energy of the Cooper pair.  These normal electrons then scatter off impurities in a boundary layer of
  thickness $\ell$=mean free path (hereafter called the ``$\ell-$layer'') and in the process transfer this momentum to the body as a whole. 
  This has to happen with  no dissipation of energy and  no increase in entropy. In the following we will discuss how it may or may
  not happen and experimental implications. We assume temperature is sufficiently low that impurity scattering dominates
  over phonon scattering.

  Note  the role of  the Faraday field in Fig. 5. It accelerates the supercurrent in the S region within 
  $\lambda_L$ of  the boundary and it
 transfers momentum to the ions in that region in clockwise direction, opposite to the motion of the body. The total  momentum that needs to be
 transferred to the body by impurity scattering is larger by a factor $R/(3\lambda_L)$
  than the net  momentum acquired by the body (Eq. (47)) because it has to compensate
 the momentum in opposite direction transferred by the Faraday field.

 A slightly different scenario for the momentum transfer within the conventional theory is the following \cite{tinkham}.
The supercurrent is carried by Cooper pairs, all having the same center of mass
  momentum $(q,0,0)$ parallel to the phase boundary. This is equivalent to a rigid shift of the
  `smeared' Fermi distribution (smeared by the superconducting energy gap $\Delta$). In a free electron
  approximation, a member of a Cooper pair at the leading edge of the distribution has maximum
  kinetic energy $(\hbar^2/2m_e)(k_F+q/2)^2$, with $k_F$ the
  Fermi wavevector. If this electron scatters to the empty state $(-k_F+q/2,0,0)$ the energy decrease is
  $\hbar^2 k_F q/m_e$, and equating this to the energy cost to break a pair, $2\Delta$,
  gives the criterion for the critical momentum $\hbar q_c$ and the critical velocity  within BCS theory \cite{tinkham}:
  \bmath
  \beq
  v_c=\frac{\hbar q_c}{2m_e}=\frac{\Delta}{\hbar k_F} .
  \eeq
  On the other hand, within London theory the supercurrent velocity when a magnetic field $H$ is applied is
  \beq
  v_s=\frac{e}{m_e c}\lambda_L H .
  \eeq
  \emath
  For $H=H_c$, Eqs. (50a) and (50b) are the same within a numerical factor of order 1 \cite{tinkham}.
 In this scenario the supercurrent stopping occurs through scattering of
 Cooper pair members (off impurities or defects or
  phonons)  and in the process breaking up the Cooper pairs with energy gain exceeding $2\Delta$. These scattering processes
  will also transfer momentum from the supercurrent to the body. Both this scenario and the one discussed 
  previously \cite{eilen,halperin}
 lead to the same contradictions and  we will 
  focus on the  one described earlier, depicted in Fig. 5,  in what follows.
  
     \begin{figure}
   \resizebox{8.5cm}{!}{\includegraphics[width=6cm]{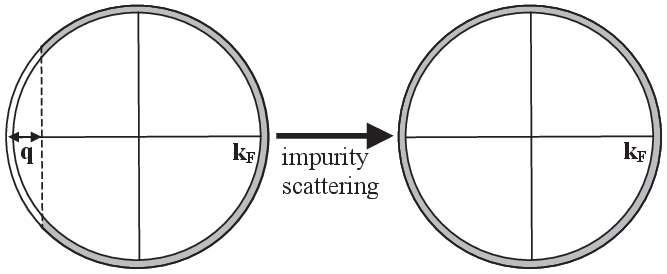}}
 \caption { The grey sliver of thickness $(\dot{r}_0\tau/\ell)k_F$ denotes the states in the region of width
 $\ell$ next to the phase boundary that become occupied by the normal  electrons 
 that originated in the dissociation of Cooper pairs with center of mass
 momentum $q$ in the horizontal ($x$) direction in time $\tau$. In the left panel, states with  $-k_F\leq k_x\leq -k_F+q$ 
 are unoccupied. Impurity scattering randomizes the occupation of states in the sliver making it isotropic (right panel)
 as the momentum gets transferred to the body as a whole.}
 \label{figure1}
 \end{figure}

  \section{phase space distribution}
  The mechanical momentum of a Cooper pair is $p_{mech}=2m_e v_s=2e\lambda_L H_c/c \equiv \hbar q$. Consider a Cooper pair with electrons of wavevector $(-k_x+q/2,-k_y), (k_x+q/2,k_y)$. 
  We assume the directions $x$ and $y$ are parallel and perpendicular to the phase boundary respectively.
  When the pair dissociates, the resulting electrons need to occupy states within $k_BT$ of the Fermi energy.
  Assuming the temperature is very low, the resulting electrons will be essentially on the Fermi surface.
  If we assume that momentum is conserved when the pair dissociates, there is essentially {\it a single state} on the Fermi
  surface  into which the electrons
  can dissociate, namely $(q/2,\sqrt{k_F^2-q^2/4})$, $(q/2,-\sqrt{k_F^2-q^2/4})$, with $k_F$ the Fermi wavevector. It is clear that the transition cannot take place if 
  all Cooper pairs have to dissociate into a single electronic state.
  
  However, it is argued \cite{halperin} that in this process momentum in the $y$ direction does not have to be conserved,
  rather the momentum difference in direction perpendicular to the phase boundary can be picked up by the superfluid \cite{halperin}.   If so, the resulting normal states for the electrons will be of the form
  $(-k_x'+q/2,k_{y1}), (k_x'+q/2,k_{y2})$ with $(k_x'-q/2)^2+k_{y1}^2=(k_x'+q/2)^2+k_{y2}^2=k_F^2$. 
  Assuming these states are at the Fermi surface, the left panel in Figure 6 shows the region that they occupy.
  The grey sliver of thickness $\Delta k=(\dot{r}_0 \tau/\ell)k_F$ accommodates the electrons originating from dissociation of
  Cooper pairs in the time interval $\tau$ assuming they lost their extra kinetic energy in the process. Note that $q>>\Delta k$.
  States in the arc to the left of the dashed vertical line on the left panel, that have $-k_F\leq k_x\leq -k_F+q$, are excluded, because their `partner' with $k_x'=k_x+q$ would necessarily
  be outside the Fermi surface. Then, this anisotropic distribution would relax through impurity scattering to the equilibrium distribution
  shown on the right panel in Fig. 6, transferring the momentum to the body. Here for simplicity we are ignoring the effect of
  electron-electron interactions; their possible role will be discussed in a later section.

 How does finite temperature affect this argument? The products of the dissociating Cooper pair can have energies within 
 $k_BT$ of the Fermi energy. So their maximum wavevector in the $x$ direction is
 $k_x^{max}=k_F+\delta k$, with
 \beq
 \delta k=\frac{T}{T_F}k_F
 \eeq
 with $T_F$ the Fermi temperature. On the other hand, the center of mass momentum of the Cooper pairs is
 \beq
 q=2\frac{e\lambda_L}{\hbar c}H_c \sim \frac{1}{2\lambda_L}.
 \eeq
 With a typical $\lambda_L \sim 500 \AA$, $k_F\sim 1\AA^{-1}$, 
 temperature $T\sim 1K$ or below and $T_F \sim 50,000K$,  it is clear that
$ \delta k<<q$, so we can disregard the effect of temperature in this argument.
 
 It is obvious  from Fig. 6 that the distribution on the left panel is more `ordered' than that on the right panel, 
 hence that impurity scattering generates entropy. Let us  estimate this entropy increase quantitatively.
 For simplicity we assume the system is at a very low temperature $T_0$. 
 The solid angle corresponding to the excluded states in Fig. 6 is $2\pi q/k_F$. If $\Delta N$ is the number of electrons
 resulting from dissociation in time $\tau$, $\Delta N_1=(q/(2k_F)\Delta N$ is the number of single particle states to the
 left of the dashed line in Fig. 6  that are `forbidden' by momentum conservation. The entropy increase in going from the left to the right panel in 
 Fig. 6 is
 \beq
 \Delta S=k_B ln\Omega
 \eeq
 where $\Omega$ is the number of ways to distribute $\Delta N_1$ indistinguishable particles in $\Delta N$ boxes. This yields
 \beq
 \Delta S=k_B\Delta N\frac{q}{2k_F}(ln(\frac{2k_F}{q})+1) .
 \eeq
 With $q/(2k_F)\sim 1/1000$, 
 \beq
 \Delta S \sim 0.01 k_B \Delta N
 \eeq
 or $0.01 k_B$ per electron that came from dissociation of a Cooper pair. This is an enormous increase in entropy.
 
 The argument above assumes that the system is at a sufficiently low  temperature $T_0$ that there are much fewer
 than $\Delta N$ thermally excited electrons and holes within $k_B T_0$ of the Fermi energy. 
 This requires $k_BT_0<(\dot{r}_0/\ell)\epsilon_F$, admittedly a very low temperature.
 Nevertheless, the important conclusion is that in this regime at least the entropy increase
 $per$ $electron$ is given by Eq. (54).
 This violates the principle that in a reversible thermodynamic process the increase in entropy should be
 arbitrarily small provided the transition occurs sufficiently slowly. Eq. (54) is incompatible with
 the superconductor-normal transition being reversible.

\section{reversibility and the Meissner transition}
           \begin{figure}
 \resizebox{8.5cm}{!}{\includegraphics[width=6cm]{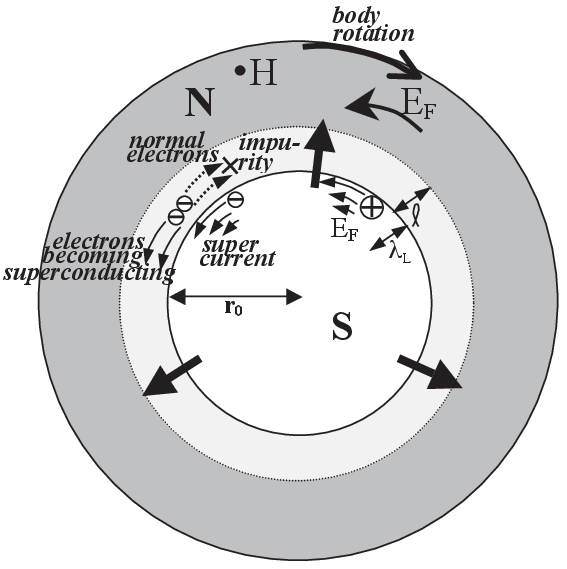}}
 \caption { N-S transition in a magnetic field (Meissner effect). The phase boundary moves outward.
 According to the conventional theory, normal electrons next to the phase boundary acquire the momentum of the supercurrent
 and leave behind normal electrons moving in the opposite direction (dotted arrows) that scatter off impurities and transfer
 the momentum to the body.
 }
 \label{figure1}
 \end{figure} 
  
  We next discuss how the conventional theory envisions the Meissner effect,  the process where a normal
  metal in a magnetic field expels the magnetic field and becomes superconducting. Figure 7 shows 
  the schematics, i.e. the inverse process to the one shown in   Fig. 5.
  
  In this case, the role of the Faraday field $E_F$ in the S region close to the phase boundary is to $decelerate$
  the electrons in the supercurrent as they move deeper into the S region (by way of the phase boundary moving outward),
  and in the same region (within $\lambda_L$ of the phase boundary)  impart counterclockwise momentum to the ions. The actual rotation of the body is clockwise,
  and the electrons joining the supercurrent acquire counterclockwise momentum in direction opposite to the force exerted on
  them by the Faraday electric field. How does the conventional theory explain these processes?
  
  According to the conventional theory \cite{eilen,halperin} as normal electrons next to the phase boundary
  form Cooper pairs and join the supercurrent, they `spontaneously' acquire the momentum of the supercurrent.
  To the best of this author's knowledge the dynamics of this process is not explained further. As this process happens,
  a momentum imbalance is created in the normal region, i.e. the normal electron distribution  is left with  equal
  momentum in the opposite direction (dotted arrows in Fig. 7). The normal electron distribution then relaxes
  through impurity scattering transfering this  momentum to the body as a 
  whole, generating the body's rotation.

           \begin{figure}
 \resizebox{8.5cm}{!}{\includegraphics[width=6cm]{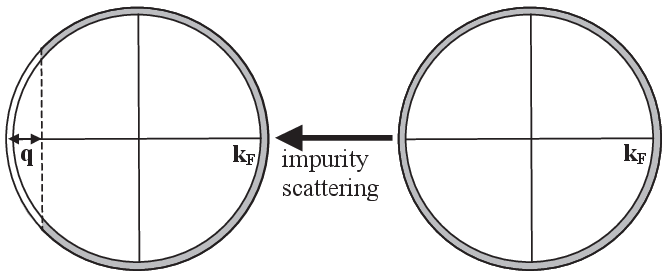}}
 \caption { The reverse process of Fig. 6. If the S-N transition as described by the conventional theory of 
 superconductivity was truly reversible, in the N-S transition normal electrons isotropically distributed (right panel) would scatter off impurities
 and end up with momentum states with $-k_F\leq k_x\leq -k_F+q$  unoccupied as depicted on the left panel.
 Clearly that cannot happen, which means that  the process is not reversible.
 }
 \label{figure1}
 \end{figure} 
 
  However, we  point out that the process just described, depicted in Fig. 7, is $not$ the reverse of the process
  depicted in Fig. 5. Reversibility means reversing the arrow of time.  In the S-N transition $first$ the Cooper pair dissociates into normal electrons inheriting the momentum of the
  supercurrent. $Then$, these normal electrons lose their extra momentum by impurity scattering, over a time period
  $\tau$. So the processes occur in sequence over a time period $\tau$. Therefore one cannot say that the reverse process 
  is that  first electrons join the supercurrent and then
  normal electron scatter, as depicted in Fig. 7. Rather, the reverse process of Fig. 5 would be that
  $first$ normal electrons in the $\ell-$layer incident on impurities from random directions would scatter preferentially with momentum
in the direction of the momentum of the supercurrent, $then$ they would  form Cooper pairs and join the supercurrent.
The first stage would correspond to reversing the process shown in Fig. 6, shown schematically in Fig. 8. 
Clearly  that cannot physically happens because it would imply lowering the entropy of the universe.

  This illustrates again that the physical processes invoked in the conventional
  theory to explain the S-N and N-S transitions  \cite{halperin} are $not$ reversible.
  
  \section{temperature dependence of the transition speed}
  
Let us examine how the speed of the transition is affected by temperature within the conventional understanding of
  superconductivity. According to the conventional theory, the processes described in the previous section involve
  Andreev reflection \cite{halperin}. In such processes when a normal electron is incident on the N-S phase boundary, it 
  is reflected as a hole and a Cooper pair is created. In order for this to have a non-negligible probability to
  happen, both the incident electron and the reflected hole have to be within $k_BT$ of the Fermi energy.
  This implies that at low temperatures only a small fraction of normal electrons can participate in this
  process, and this fraction will go to zero as $T$ approaches zero. In other words, the theory predicts that
  for temperatures sufficiently close to zero the Meissner effect will not take place, or will take an arbitrarily long time.
  
  Let us estimate the time involved. In the normal layer of thickness $\ell$ next to the phase boundary ($ \ell$-layer) a fraction $k_BT/\epsilon_F$ of electrons can potentially undergo Andreev scattering at the interface and
  become superconducting. The process also involves relaxing the resulting
  momentum imbalance of the remaining normal electrons through impurity scattering, which takes time $\tau$. Therefore
  at a minimum the time required to convert all the normal electrons in the $\ell-$layer to superconducting electrons carrying a current will be
  \beq
  t_\tau=\tau\frac{\epsilon_F}{k_BT}.
  \eeq
  This implies that the speed of motion of the phase boundary cannot exceed
  \beq
  \dot{r}_0^{max}=\frac{\ell}{t_\tau}
 =\frac{v_F k_BT}{\epsilon_F}
 \eeq
  or equivalently
   \beq
  \dot{r}_0^{max}=\frac{5.1}{\epsilon_F(eV)^{1/2}}T(mK)\frac{cm}{s} .
  \eeq
  For example, for $\epsilon_F=10eV$ and $T=1mK$, $\dot{r}_0^{max}=1.6 cm/s$.
  For a cylinder of radius $R=1cm$ at $T=10 \mu K$, the transition cannot occur in less than $160$ seconds.
  
  The speed of the transition is also limited by Faraday's law, as discussed in Sect. III. As the temperature decreases
  the normal state conductivity $\sigma$ increases and that implies that the time for the transition as described in sect. III
  will increase as given by Eq.(33). However at very low temperatures $\sigma$ will reach a constant value and
  the speed of the transition as limited by Faraday's law will reach a limiting value. Instead, the physics discussed in 
  this section implies that the time for the transition will increase without bounds as the temperature is lowered further.
  Reversibility implies that the transition time for the S-N transition should also increase without bounds in this situation.
  
  We know of no experimental evidence suggesting that the Meissner transition (and its reverse) will 
  freeze out as the temperature becomes very low. By increasing the disorder in the sample we can decrease the time of the
  transition allowed by Faraday's law Eq. (33) so that the physics discussed here will dominate.
  We suggest this should be probed experimentally.
  If it is found that the transition does not slow down significantly as the temperature is lowered
    it will call into question the validity of the scenario required by the conventional theory.

  \section{the question of equilibrium}
  
How does the conventional theory envision the equilibrium along the normal-superconductor
  phase boundary in the presence of a magnetic field? The problem was first considered by 
  H. London in 1935 \cite{londonh}, however the issues discussed here were not raised.
  
  According to the conventional theory, ``in thermal equilibrium pair recombination and pair breaking processes
  occur all the time'', ``there is a detailed balance of pair decay and pair creation'' \cite{scalapino}.
  ``The process and its reverse will be happening all the time in thermal equilibrium and there is clearly no
  generation of entropy in this situation'', ``the process of incident electron being reflected as holes will be balanced by the process of incident holes being
  reflected as electrons'' \cite{halperin}.
  
  However, according to the discussion in Sect. VI, when Cooper pairs become normal their momentum distribution 
  is not isotropic as depicted in Fig. 6, and becomes isotropic through impurity scattering, in the process generating
  entropy. In the reverse process involving Andreev reflection, relaxation of the normal electron momentum by
  impurity scattering also generates entropy. Therefore, contrary to the statements above \cite{halperin,scalapino} we argue that the
  conventional scenario predicts that {\it in thermodynamic equilibrium there is in fact continuous generation of entropy}
  from  these processes..
  Clearly this does not make sense and implies that the conventional theory is $incompatible$ with a 
  situation of thermodynamic equilibrium between normal and superconducting phases in a magnetic field. We believe that
  ample experimental evidence exists that such equilibrium does exist in nature without continuous generation
  of entropy.

  \section{ the role of electron-electron interaction}
  
  In the foregoing we have assumed for simplicity that electrons don't interact with each other. In fact they do. 
  How will this modify the arguments presented earlier? 
  
  Electron-electron scattering will tend to make the anisotropic distribution in Fig. 6 isotropic but not relax
  the momentum, since it conserves total momentum. In the absence of impurity scattering the momentum
  distribution would change as shown in Fig. 9, giving rise to a slightly shifted isotropic
  Fermi distribution.  In time $\tau$ the phase boundary moves a distance $\dot{r}_0\tau$. 
  Once the distribution becomes isotropic  the electrons
  that became normal in time $\tau$ will have
  shared their extra momentum with all the electrons in the $\ell-$layer, resulting in a small 
shift $\delta q$ in the Fermi surface as shown in the right panel of Fig. 9, given by  $\delta q=((q/2)\dot{r}_0\tau/\ell=(q/2)\dot{r}_0/v_F$. Then, this slightly shifted Fermi sea will relax by impurity scattering transferring its momentum to the lattice.
  
 The time scale for electron scattering is $\tau_{ee} \sim \hbar \epsilon_F / (k_B T)^2$, which at low temperatures
  is usually larger than the impurity scattering time $\tau$, so impurity scattering will dominate the relaxation. 
  Nevertheless, let us assume a very clean sample where a combination of electron-electron scattering and
  a small amount of impurity scattering will
  first randomize the momenta as shown in Fig. 9 conserving total momentum, and  subsequently impurity scattering will
  transfer the momentum to the body as a whole. The first process will generate entropy as given by Eq. (54),
  and the second process will generate additional entropy, as discussed in what follows.
  
  \section{the conventional argument}
  
The conventional argument regarding entropy generation 
  in the process of transfering momentum goes as follows \cite{halperin}. The extra momentum
  carried by each dissociating electron $q/2$ is shared by the entire Fermi sea in the $\ell-$layer, giving
  a very small momentum shift
   \beq
   \delta q=((q/2)\dot{r}_0\tau/\ell=(q/2)\dot{r}_0/v_F 
   \eeq
    to the entire Fermi sea. Some
  entropy will be generated as the momentum is transferred to the body through impurity scattering, but it is argued that it is quantitatively small, goes to zero
  as the speed of the transition goes to zero, and  may not be inconsistent with observations. Let us analyze it
  quantitatively.

         \begin{figure}
 \resizebox{8.5cm}{!}{\includegraphics[width=6cm]{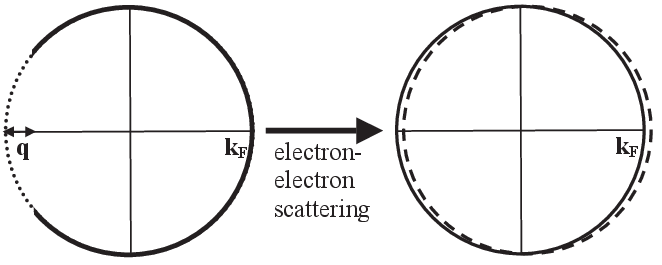}}
 \caption { If electron-electron scattering dominates, the distribution will become isotropic conserving total 
 momentum, so the Fermi surface ends up shifted slightly to the right by an amount
 $\delta q=(\dot{r}_0\tau/\ell)q/2$ (dashed line). }
 \label{figure1}
 \end{figure} 
 
  The mechanical momentum density in the $\ell$-layer under this assumption  will be
  \beq
 \mathcal{P}_q= n_s \frac{\dot{r}_0 \tau}{\ell}\frac{\hbar q}{2}= n_s \frac{\dot{r}_0}{v_F}\frac{\hbar q}{2}= n_s \frac{\dot{r}_0}{v_F}\frac{e\lambda_L}{c}H_c
 \eeq
 that was acquired from the $(n_s/2) \dot{r}_0 \tau (2\pi r_0h) $ Cooper pairs of center of mass momentum $\hbar q$ that dissociated in time $\tau$. The mechanical momentum density Eq. (60) is 
 equivalent to an electric current density  $J_q$ given by \cite{rostoker} 
 \beq
 J_q=\frac{e}{m_e}  \mathcal{P}_q=\frac{e}{m_e}\frac{n_s}{2}\frac{\dot{r}_0 \tau}{\ell}\hbar q=\frac{n_s e^2}{m_e} \frac{\dot{r}_0}{v_F} \frac{\lambda_L}{c}H_c.
 \eeq
 When this current decays by collisions with impurities, its mechanical momentum is transferred to the body
 as a whole. 
  The momentum density Eq. (60)  has
 associated with it a kinetic energy density
  \beq
 K_q=\frac{\mathcal{P}_q^2}{2m_e n_s}=
 \frac{m_e}{2e^2 n_s} J_q^2=\frac{H_c^2}{8\pi}(\frac{\dot{r}_0}{v_F})^2
 \eeq
 and the decay of this current will generate  entropy per unit volume   $K_q/T$.

On the other hand the Faraday electric field Eq. (31) that exists in this region also distorts slightly
the Fermi distribution, giving rise to a normal current 
\beq
J_F=\sigma E_F=\frac{n  e^2}{m_e} \frac{\dot{r}_0 }{v_F} \frac{\ell}{c}H_c=\frac{\ell}{\lambda_L}
\frac{n}{n_s}J_q
\eeq
where we have assumed Eq. (30) for the normal state conductivity. We will assume   $n=n_s$ for simplicity (in general $n$ can be larger than $n_s$).
Note that the
currents $J_q$ and $J_F$ are of similar form and similar magnitude for the case 
$\ell \sim \lambda_L$, $n\sim n_s$, and they both become vanishingly small as the speed of motion of
the phase boundary $\dot{r}_0$ becomes vanishingly small.
The kinetic energy density associated with $J_F$ will be similar to $K_q$.
However, a difference between $J_q$ and $J_F$  is that $J_F$ flows over the entire normal region $r_0(t)\leq r \leq R$
while $J_q$ only flows in the region $r_0(t)\leq r \leq r_0(t)+\ell$. 
As a consequence, the total entropy generated by the momentum-transfer current $J_q$ will be much smaller than the 
entropy generated by the Faraday current 
$J_F$. Note also that the momentum transferred from $J_F$ to the body by collisions is cancelled by the opposite momentum
transferred directly by $E_F$ to the ions.

In the layer of thickness $\ell$ adjacent to the phase boundary  the total current density is then
\beq
J_{tot}=J_q+J_F .
\eeq
We have already considered the effect of $J_F$ in producing entropy in  section IV, so we need to substract its contribution here to avoid double
counting. The net additional contribution of the current density  $J_{tot}$ is then  a kinetic energy density
 \beq
 K_{excess}=\frac{m_e}{2e^2 n_s}(J_{tot}^2-J_F^2)=\frac{H_c^2}{8\pi}(\frac{\dot{r}_0}{v_F})^2(1+\frac{2\ell}{\lambda_L})
 \eeq
 where we have used Eq.  (38).
 This excess kinetic energy resides in the region $r_0 \leq r \leq r_0+\ell$ and was generated
 by motion of the phase boundary in time $\tau$ and associated conversion of the
 supercurrent into normal currrent. As this excess normal current decays through normal scattering processes, the entropy
 per unit volume  generated 
 for the cylinder  of radius $R$ in the process of becoming normal due to this physics will be, from Eq. (65)
 \beq
 \Delta S_{c}=\frac{H_c^2}{8\pi T} \frac{1}{\pi R^2}\frac{1}{\tau}(1+\frac{2\ell}{\lambda_L}
) \int_0^{t_0} dt (\frac{\dot{r}_0}{v_F})^2(2\pi r_o \ell)
 \eeq
 and changing integration variables from $t$ to $r_0$ and using Eq. (35)
  \beq
 \Delta S_{c}=\frac{H_c^2}{4\pi T}p\frac{\lambda_L}{\ell}\frac{\lambda_L}{R}(1+\frac{2\ell}{\lambda_L}
 ) ) \int_0^{1-\delta} dx\frac{1}{ln(\frac{1}{x})}  .
 \eeq
 In the upper limit of the integral Eq. (67) we have included a small cutoff parameter $\delta>0$ to render the
 integral non-divergent. The weak divergence for $\delta\equiv0$ originates in the fact that the speed of the phase boundary $\dot{r}_0$,
 Eq. (35), diverges as $r_0\rightarrow R$. This is of course unphysical and stems from the
 approximation inherent in Eq. (35) that  the only factor limiting the speed of the
 phase boundary is Faraday's law \cite{pippard}. In reality, as the speed becomes large other factors will set in to limit $\dot{r}_0$, e.g. finite thermal conductivity. This has been discussed by Faber \cite{faber}, who also showed experimentally
 that Eq. (35) for $\dot{r}_0$ is accurately satisfied over the entire range of $r_0$ except for $(R-r_0)/R \lesssim 0.05$  
 where it is cut off.
 
 Thus, the integral in Eq. (67) is of order unity, and we have   approximately
   \beq
 \Delta S_{c}= \frac{H_c^2}{4\pi T}p\frac{\lambda_L}{\ell}\frac{\lambda_L}{R}(1+\frac{2\ell}{\lambda_L}
) )  .
 \eeq
This  is smaller than the increase in entropy per unit volume due to Joule heat generation from eddy currents 
\beq
\Delta S_{Joule}=\frac{H_c^2}{4\pi T}p
\eeq
by a
factor $\sim (\lambda_L/R)$ if we assume $\lambda_L\sim\ell$.

     \begin{figure}
 \resizebox{6.5cm}{!}{\includegraphics[width=6cm]{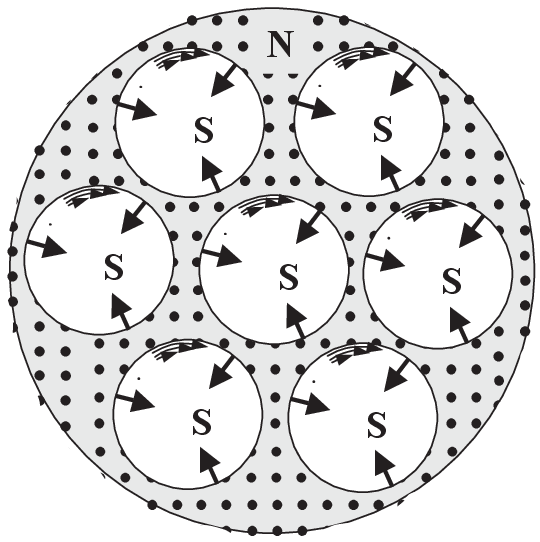}}
 \caption {Superconducting cylinder becoming normal through a route where it first breaks up into many superconducting domains
 of initial radius $R_0$.  }
 \label{figure1}
 \end{figure}

It should be noted that  there is an important difference between the entropy contributions Eqs. (68) and (69). The entropy contribution from eddy currents
 Eq. (69) does not depend on the particular way the sample went normal, since eddy currents are generated everywhere in the normal parts of the sample.
 Assume that instead of a process where the boundary moves uniformly inward as in Fig. 4, the sample goes normal through a kind of 
 `intermediate state' where the magnetic field breaks up the superconducting regions into domains that evolve separately,
 as shown in Fig. 10. Each superconducting region in Fig. 10 has a boundary supercurrent density given by Eq. (39). If the regions shown in Fig. 10 have initial radius $R_0$, the change in entropy per unit volume due to the
 supercurrent stopping  is given by 
 \beq
\Delta S_{c}=\frac{H_c^2}{4\pi T} p(\frac{\lambda_L}{\ell})(\frac{\lambda_L}{R_0}) (1+\frac{2\ell}{\lambda_L}
)  
\eeq
instead of by Eq. (68). 
Instead, the entropy contribution from eddy currents is still given by Eq. (69) that does not depend
on the size of the domains. Thus, the extra entropy generation from the supercurrent stopping
will be much larger if the transition proceeds through the route shown schematically in Fig. 10 and $R_0$ is small.  
This is because there is a lot more supercurrent in the situation shown in Fig. 10 than there is in the case shown in Fig. 4.

Nevertheless, it is argued that the increase in entropy obtained here is sufficiently small that it can be ignored,
or perhaps that it is cancelled by other small effects that have been neglected \cite{halperin}. However, in the next section we will
argue that the Joule heat and resulting entropy generated by the processes described here is in fact much larger.

\section{the flaw in the conventional argument}

Here we will show that the Joule heat generated by the stopping of the normal current is in fact much larger than
calculated in the previous section. In addition to causing problems with the second law of thermodynamics,
this violates either the first law of thermodynamics or the essential assumption of the conventional theory of superconductivity
that the kinetic energy of the supercurrent equals the difference in free energies in the normal and
superconducting states.

The point we wish to make is that the kinetic energy associated with the normal current that originates in the
decay of the Cooper pairs is in fact much $larger$ than Eq. (62). The reason is, as discussed in Sect. VII,
that a fraction $q/(2k_F)$ of the states available to the dissociating electrons (left panel of Fig. 6) are 'blocked' by momentum
conservation, because if one electron of the dissociating pair occupies those states the other electron would have
energy higher than the Fermi energy.  
Quantitatively, this implies that the Fermi level in the $\ell-$layer gets shifted to
\beq
\bar{k}_F=k_F(1+\frac{1}{3}\frac{q}{2k_F}\frac{\dot{r}_0\tau}{\ell})
\eeq
as Cooper pairs dissociate into normal electrons. This gives an extra kinetic energy per particle
\beq
\Delta \epsilon_{kin}=\frac{\hbar^2}{6m_e} k_F q \frac{\dot{r}_0\tau}{\ell} .
\eeq
for the electrons coming from dissociating Cooper pairs.
To obtain   the extra kinetic energy in the $\ell-layer$ we multiply Eq. (72) by $n_s\dot{r}_0\tau/\ell$ and obtain
\beq
\bar{K}_q=\frac{H_c^2}{8\pi}(\frac{\dot{r}_0}{v_F})^2(\frac{4k_F}{3q})=K_q( \frac{4k_F}{3q})
\eeq
which is much larger than the earlier result Eq. (62).

The argument just presented is approximate because it assumes that the density of states at the Fermi surface for
the dissociating electrons is constant, and this is not the case for $q\neq 0$. Nevertheless, it is a good approximation.
We have verified this by numerical calculations in both two- and three-dimensional geometries.
In a 2-dimensional geometry,  Eq. (72) is modified as
\beq
\Delta \epsilon_{kin}=\frac{\hbar^2}{2m_e \pi} cos^{-1}(1-\frac{q}{k_F})\frac{\dot{r}_0\tau}{\ell} \sim
 \frac{\hbar^2}{2m_e\pi} \sqrt{\frac{2q}{k_F}}\frac{\dot{r}_0\tau}{\ell}
\eeq

          \begin{figure}
 \resizebox{8.5cm}{!}{\includegraphics[width=6cm]{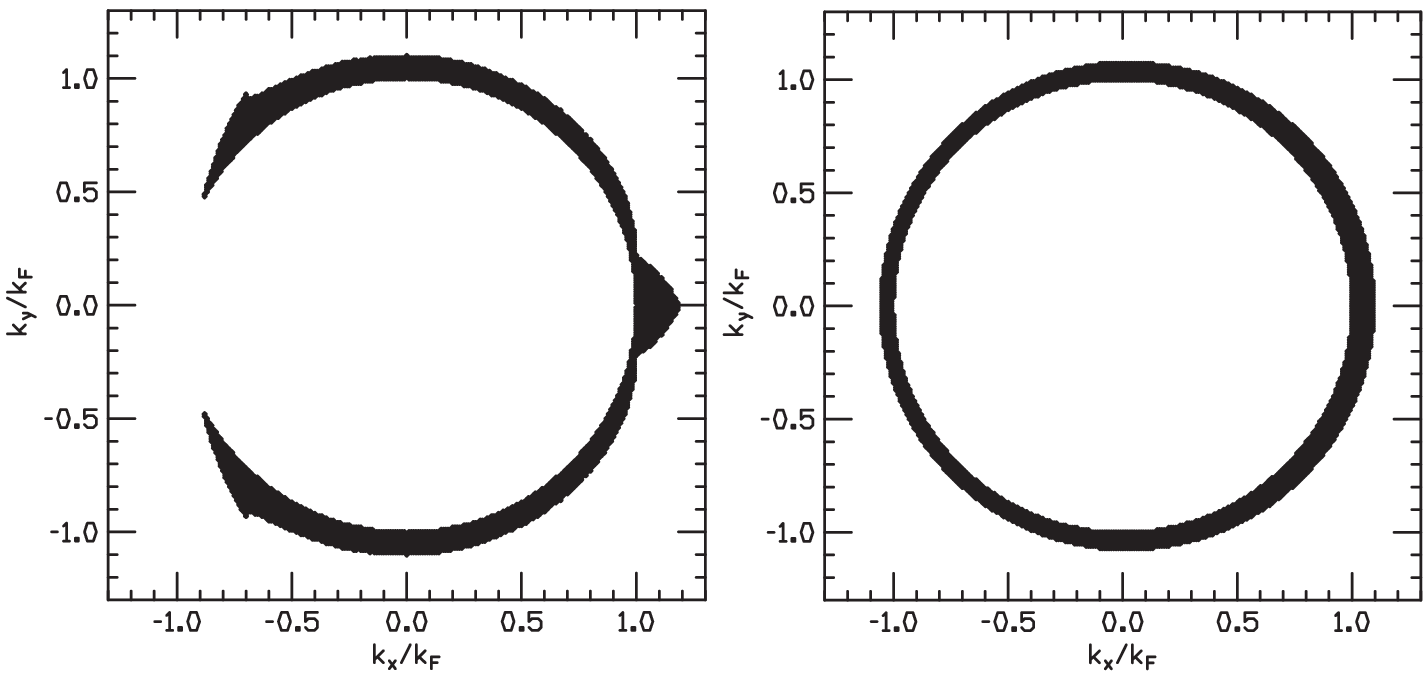}}
 \caption {Example of momentum distribution of electrons from dissociating Cooper pairs 
 (left panel). We show an unrealistic example to illustrate the qualitative behavior.
 2 dimensions, $q/k_F=0.3$, fraction of particles added to the Fermi sea $=0.16$. The right panel shows the
 isotropic distribution with  shift $\delta q=0.16 (q/2)$, as assumed in Eq. (59). 
 The ratio of kinetic energies of the left and right panels is 8.9
 }
 \label{figure1}
 \end{figure}

As an example,
Fig. 11 shows numerical results for 2 dimensions, $q/k_F=0.3$, $\dot{r}_0\tau/\ell=0.16$.
For these extreme parameters the analytic formula Eq. (74) is not very accurate, since it assumes the
distribution is uniform except for the 'hole' in the back end, which clearly is not the case for the left panel of Fig. 11. 
The ratio of kinetic energies in the left and right panels of Fig. 11 from the numerical calculation is 8.9 while the analytic formulas Eqs. (74) and (59) yield 17.9 in this case. 
As somewhat more realistic examples, in 3 dimensions with $\dot{r}_0\tau/\ell=0.035$, with $q/k_F=0.2$ and $q/k_F=0.1$ the ratios of kinetic
energies from the numerical calculations are $16.3$ and $5$, while the analytic formulas Eq. (72) and (59) predict 
$13.3$ and $6.67$, i.e. $4k_F/3q$. In a realistic situation $4k_F/3q\sim 1000$ or larger.

Returning to Eq. (73), we conclude then that the dissipation of Joule heat from this kinetic energy will far exceed the
Joule heat dissipated due to the Faraday field in the $\ell-$layer. Ignoring the Faraday field
in the $\ell-$layer we then obtain
by time  integration following the steps after Eq. (65) and using $q=1/(2\lambda_L)$
\beq
\bar{K}_q^{total}=\frac{H_c^2}{8\pi} \frac{16}{3}p \frac{\lambda_L}{\ell}\frac{\lambda_L^2k_F}{R}
\eeq
for the Joule heat per unit volume {\it in addition} to the Joule heat Eq. (26) dissipated due to the Faraday field.

The fact that the S-N transition is governed by thermodynamics requires that when the 
supercurrent stops its entire kinetic energy is used up in paying for the free energy cost of
becoming normal, $H_c^2/8\pi$ per unit volume. This follows from the fact that the kinetic energy associated with the superfluid current Eq. (39) is
\beq
K_s=\frac{m_e}{2e^2n_s}J_s^2=\frac{m_e}{2e^2n_s}  (\frac{n_s e^2\lambda_L} {m_ec}H_c)^2 =\frac{H_c^2}{8\pi}
\eeq where we have used Eq. (38). If this was not the case, there could
not be phase equilibrium between superconducting and normal phases in the
presence of a magnetic field \cite{londonh}. 
This argument indicates that in fact the kinetic energy $\bar{K}_q^{total}$ 
has to be exactly  zero in order to not violate  energy conservation.

Instead, Eq. (75) is quantitatively very significant. For example, for $p=0.1875$, $k_F=1\AA^{-1}$ and
$\lambda_L=500\AA$, it says that for $R=0.1mm$ this extra Joule heat is $1/4$ of the total
kinetic energy of the supercurrent. This will also be the case for a larger sample with domains of
radius $R_0=0.1mm$ as shown in Fig. 10. This violates energy conservation because
there is no source for this energy. It also violates the second law, since it predicts an
entropy generation $K_q^{total}/T$ which is comparable to the entropy generated by eddy currents
Eq. (27) which we have shown  in Sect. IV accounts for the entire entropy generated in the transition
according to thermodynamics.

\section{recapitulation and consistency check}
Let us rederive the results of the previous section in a simpler way to check their consistency.

The kinetic energy of the supercurrent amounts to each electron having an extra kinetic energy
\beq
\epsilon_s=\frac{\hbar^2q^2}{8m_e} .
\eeq
All that energy should be paid to dissociate a Cooper pair according to Eq. (76). However, it is impossible that 
electrons becoming normal keep the momentum of the supercurrent but none of its kinetic energy. The conventional
argument \cite{halperin} discussed in Sect. XII says that each unpairing electron shares its momentum with all the 
electrons in the $\ell-$layer, and to do that it only needs a tiny amount of kinetic energy, namely
\beq
\epsilon_{conv}=\frac{\hbar^2q^2}{8m_e}( \frac{\dot{r}_0\tau}{\ell})=\epsilon_s ( \frac{\dot{r}_0\tau}{\ell})^2
\eeq
so the difference between Eqs. (77) and (78) is available to unpair the Cooper pair. This already violates conservation
of energy, albeit by a small amount.

However, we have shown in the previous section that in fact Eq. (78) is incorrect. Because momentum in direction 
parallel to the interface has to be conserved when a Cooper pair dissociates, the kinetic energy of each
dissociating electron is
\beq
\epsilon_{correct}=\frac{\hbar^2q^2}{8m_e} (\frac{4}{3}\frac{k_F}{q})( \frac{\dot{r}_0\tau}{\ell})^2
=(\frac{4}{3}\frac{k_F}{q}) \epsilon_{conv}
\eeq
which is substantially larger than $\epsilon_{conv}$. 

From comparing Eqs. (79) and (77) we estimate that the total Joule heat dissipated per unit volume arising from the kinetic energy
Eq. (79) will be
\beq
Q_{kin}=\frac{H_c^2}{8\pi}(\frac{4}{3}\frac{k_F}{q})( \frac{\bar{\dot{r}}_0\tau}{\ell})
\eeq
where $\bar{\dot{r}}_0$ is the average speed of the phase boundary, which we obtain from 
\beq
\bar{\dot{r}}_0=\frac{R}{t_0}=\frac{4\lambda_L^2}{R\tau}p
\eeq
where we have used Eqs. (30), (33) and (38). Replacing in Eq. (80) yields 
\beq
Q_{kin}=\frac{H_c^2}{8\pi} \frac{32}{3}p \frac{\lambda_L}{\ell}\frac{\lambda_L^2k_F}{R}
\eeq
which differs by a factor of $2$ from our earlier result Eq. (75). The discrepancy arises because here
we have used an average for the phase boundary speed rather than using the exact expression Eq. (35)
and doing the time integration as in Sect. XII. Nevertheless this confirms the consistency of our approach
and in particular the validity of introducing a cutoff in the integral in Eq. (67).

 \section{reversibility again}
 
        \begin{figure}
 \resizebox{8.5cm}{!}{\includegraphics[width=6cm]{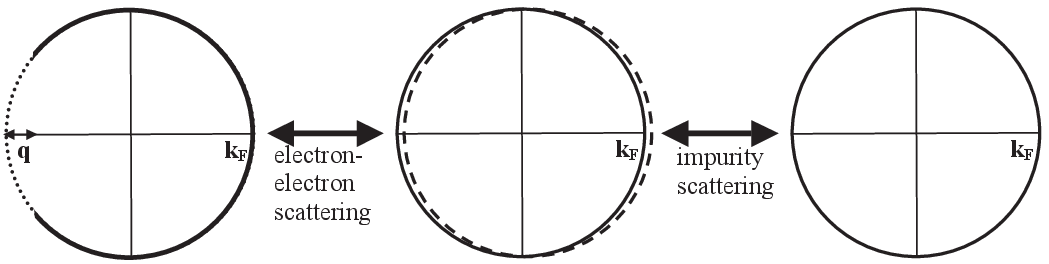}}
 \caption {Direct and reverse stages in the transition process  taking into account the effect of electron-electron interactions.  
 It is clear that these processes cannot occur from right to left. }
 \label{figure1}
 \end{figure} 
 
 Let us consider again the question of reversibility including the role of electron-electron interactions,
 under the assumption that electron-electron scattering occurs faster than electron-impurity scattering. 
 The processes at play are shown in Fig. 12. If the transition is truly reversible, these processes 
 should be able to happen in both directions shown by the arrows. 
 So starting from the rightmost panel, impurity scattering processes in an isotropic Fermi sea should give rise to an
 isotropic distribution that is slightly shifted to the right (middle panel), then electron-electron scattering processes
 should distort this distribution to the anisotropic form shown on the left panel, and finally electrons from
 this anisotropic distribution should condense into the supercurrent-carrying Cooper pairs. 
 It is obvious that these reverse physical processes are not physically possible because they lower the
 entropy of the universe and hence do not take place in nature.
 
\section{discussion}
We have analized the momentum transfer between the supercurrent and the body and   the production of entropy  when a superconductor in a magnetic field goes normal and in the reverse process within the conventional theory of superconductivity. For comparison we have also discussed the somewhat analogous process   of a liquid evaporating.

For the liquid-gas transition, our understanding raises no conflicts with known physical laws. 
 Entropy production can occur in a variety of ways:
transfer of heat across a temperature gradient, or friction of the piston with the cylinder walls, or internal friction in the gas phase. The entropy that
these processes generate is consistent with the laws of thermodynamics. For the total entropy generated 
to be infinitesimal   the process  has to occur infinitely slowly, 
  in which case it becomes exactly reversible, i.e. no change in the entropy of the universe. The converse
  is however not true, the process can occur infinitely slowly and still give rise to finite entropy generation.

In the case of the superconductor-normal transition as described by the conventional theory of superconductivity  \cite{tinkham,eilen,halperin}, we identified two separate  
sources for entropy generation: Joule heating due to eddy current
decay, generated by   the changing magnetic flux, and entropy generation in the process of the supercurrent stopping and transferring its mechanical
momentum first to normal electrons and from there to the body, and vice versa.  

We showed quantitatively that Joule heating from eddy currents accounts for the entire entropy generated in the transition according to the laws of thermodynamics.
This then implies that zero entropy can be  generated by the process of the supercurrent stopping and transferring its momentum to the body.
However, we analized  the process of momentum transfer from the supercurrent to the body
following the prescriptions of the conventional theory of superconductivity \cite{eilen,halperin} and 
found that entropy is necessarily generated in this process. We pointed out that there are two different  processes that
generate entropy. One is the randomization of the anisotropic momentum distribution (Figs. 6, 11) that
results from the pair dissociation. This entropy generation is non-zero even in the limit where the transition occurs
infinitely slowly. The second is the dissipation of the extra kinetic energy of the distributions (Figs. 6, 11) as Joule heat, as the
momentum is transferred to the body. This can be  quantitatively comparable to the Joule heat generated by
eddy currents and violates both conservation of energy and the second law of thermodynamics.

We have also argued that  according to the conventional theory
entropy would be continuously generated in a situation of thermodynamic equlibrium between the
normal and superconducting phases in the presence of a magnetic field, where pair recombination and pair breaking processes
continuously occur, as the resulting transient anisotropic normal electron momentum distribution 
resulting from these processes
(Fig. 6 left panel)  is  randomized by scattering
processes. Clearly this  does not make physical sense, since in a system in 
thermodynamic equilibrium no entropy is generated.

It is understandable that no entropy can be generated in the process of the supercurrent stopping for the following reason: initially the current and its
associated mechanical momentum is carried
by the condensate, all pairs having the same center of mass momentum. That is a single state, carrying no entropy. In the final state, the 
mechanical momentum is carried by the body as a whole, also a single state carrying no entropy.  
This implies that there can be no intermediate stage in the process of transferring the mechanical momentum from the
supercurrent to the body  that does carry entropy, because that entropy cannot
be destroyed again to reach the final state. This was already recognized long ago by Keesom \cite{keesom2}, who wrote ``{\it it is essential
that the persistent currents have been annihilated before the
material gets resistance, so that no Joule-heat is developed}.
Within the conventional theory of superconductivity   there is no mechanism to transfer the mechanical momentum of
the supercurrent $directly$ to the body as a whole wthout involving the normal electrons. 

The unphysicality of the conventional description is particularly striking for the case where the
superconducting region is $expanding$ instead of contracting, i.e. the Meissner effect, as discussed in Sect. VIII, where   the normal state Fermi sea in the layer
of thickness $\ell$ adjacent to the boundary would have to arrange itself, through collisions between electrons and impurities,  into the anisotropic
momentum distribution shown in the left panel of Fig. 8,  to then transfer it to the superconducting region as normal particles  bind to form Cooper pairs. 
There certainly is no physical way to justify this process which necessitates a spontaneous $lowering$ of the entropy of the universe..

In the next section we discuss how the theory of hole superconductivity \cite{holesc}  resolves these issues.

 \section{explanation within the theory of hole superconductivity}

          \begin{figure}
 \resizebox{8.5cm}{!}{\includegraphics[width=6cm]{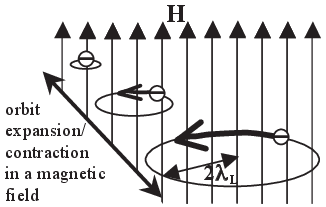}}
 \caption {When an electron expands or contracts its orbit in a perpendicular magnetic field its azimuthal velocity changes proportionally to the
 radius of the orbit due to the azimuthal Lorentz force acting on the radially outgoing or ingoing charge.}
 \label{figure1}
 \end{figure}

Here we discuss briefly how the theory of hole superconductivity resolves these issues. 
 For more details the reader is referred to references \cite{revers,disapp,momentum} and references therein.
 
 There are two questions that need to be answered. (1) how do electrons acquire or lose the momentum of the supercurrent, and (2) how does the body
 acquire or loses its momentum. The essence of the answer is that the momentum transfers are mediated by the electromagnetic field rather than
 by scattering processes, as explained  in ref. \cite{momentum}.
 
 First, how do electrons acquire or lose their momentum reversibly? This results from expansion or contraction of electronic orbits, as shown schematically
 in Fig. 13. The theory of hole superconductivity predicts \cite{sm} that when electrons become superconducting their orbits expand from a microscopic
 radius to radius $2\lambda_L$. driven by lowering of kinetic energy. When an orbit expands from a microscopic radius to radius $r$ in the
 presence of a magnetic field,  the magnetic
 Lorentz force on the outgoing electron imparts   an azimuthal speed \cite{copses}
 \beq
 v_\phi= - \frac{er}{2m_ec}H ,
 \eeq
 thus when the orbit expands to radius $r=2\lambda_L$ the azimuthal speed acquired is the speed of the Meissner current Eq. (50b). 
 Similarly when the orbit shrinks from radius $2\lambda_L$ to a microscopic radius the azimuthal Lorentz force acts in the opposite direction 
 and the supercurrent stops. 
 
 Why $2\lambda_L$ orbits? The supercurrent flowing in a surface layer of thickness $\lambda_L$ results from 
 superposition of $2\lambda_L$ orbits in the bulk, and electrons in such orbit carry orbital angular
 momentum $\hbar/2$ \cite{bohr}.
 
 The process  just described does not involve any scattering and hence is reversible.
 The reversible expansion or contraction of the orbits explains how the supercurrent starts and stops in a given external magnetic field.

 Second, how does the body as a whole acquire compensating momentum in the opposite direction? In the  process just described, momentum conservation holds because  as the orbit expands or contracts and
the azimuthal momentum of the electron changes,
a compensating azimuthal momentum is stored in the electromagnetic field, as explained in \cite{momentum}. That momentum
is retrieved by a radial flow of normal charge to compensate for the radial charge redistribution that occurs when the
electronic orbit expands or contracts. Here is where the hole-like nature of the normal state charge carriers plays a key role.

           \begin{figure} [t]
 \resizebox{16cm}{!}{\includegraphics[width=6cm]{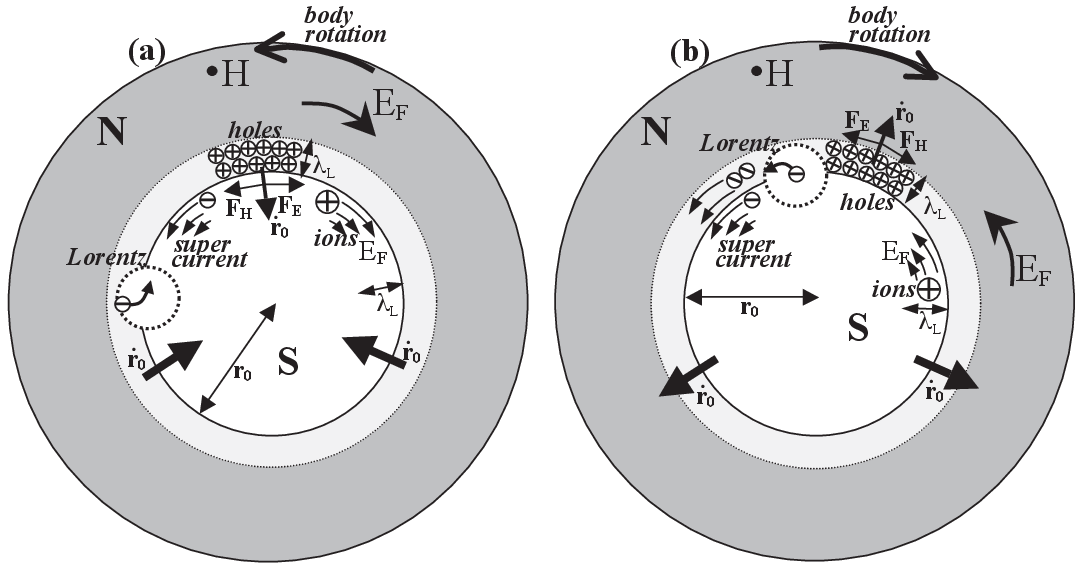}}
 \caption {(a) S-N  and  (b) N-S (Meissner effect)  transitions in a magnetic field according to the theory of hole superconductivity.
 (a) As superconducting electrons become normal their orbits (dotted circle) shrink, and the resulting Lorentz
 force stops the supercurrent, without any scattering processes. An inflow of hole carriers moving in the same direction as the
 phase boundary restores charge neutrality and transfers momentum to the body as a whole to make it
 rotate counterclockwise, without any scattering processes. (b) As normal electrons become superconducting  their orbits (dotted circle) expand, and the resulting Lorentz
 force propels the supercurrent. An outflow of hole carriers moving in the same direction as the
 phase boundary restores charge neutrality and transfers momentum to the body as a whole to make it
 rotate clockwise, without any scattering processes.}
 \label{figure1}
 \end{figure}

Figure 14 explains the physics of the processes for the S-N and N-S transitions. In the left panel (S-N transition) the superconducting region is shrinking and $2\lambda_L$ orbits 
at the boundary of the N-S region are contracting, which imparts the electrons with a clockwise azimuthal impulse,
stopping the supercurrent. At the same time, to compensate for the radial inflow of negative charge, a backflow of 
normal charge occurs: normal state hole carriers flow radially inward at speed $\dot{r}_0$, the speed of motion of the phase boundary.  The azimuthal forces on this backflow of inflowing  hole carriers 
are $F_H$ and $F_E$ shown in Fig. 14. $F_H$ is the magnetic Lorentz force  and 
$F_E$ is the electric force due to the Faraday field $E_F$, which  is given according to Faraday's law  by
\beq
E_F=\frac{\dot{r}_0}{c}H .
\eeq
Thus, $eE_F$ is precisely the magnetic Lorentz force on the hole carriers   in opposite direction to the electric force, the 
azimuthal forces on the holes exactly cancel out and the hole motion is radial.
Similarly the right panel shows the process for the N-S transition (Meissner effect). Here the orbits expand as the 
phase boundary moves out, propelling the supercurrent, and the backflow consists of holes moving radially outward together with the
phase boundary. Again, the azimuthal electric and magnetic forces on the holes are balanced.

What we have just described explains the body's rotation. The radial inflow or outflow of hole carriers
creates a torque that makes the body rotate, imparting the same angular momentum gained or lost by
the supercurrent in opposite direction, as shown quantitatively in ref.  \cite{momentum}.
It is instructive to consider the normal backflow in terms of electrons rather than holes, as shown in
Figure 15. Note that now the electric and magnetic forces on these electrons are $not$ balanced, rather
they point in the same direction. To recover force balance we need to
add another force $F_{latt}$, which is a transverse force that the periodic ionic potential exerts on 
the charge carriers moving in crossed electric and magnetic field {\it when the charge carriers are holes}.
By Newton's third law, there is an equal opposite force exerted by the electrons on the ions, which
we denote by $F_{on-latt}$. This force acts in the direction of the body rotation, it is in fact the force that makes the body rotate.
The transfer of momentum to and from the body is $not$ mediated by impurity scattering. Rather, it is mediated by the
coherent interaction of electrons near the top of energy bands, that have a negative
effective mass, with the periodic potential of the ions \cite{disapp}. 

             \begin{figure}[h]
 \resizebox{16cm}{!}{\includegraphics[width=6cm]{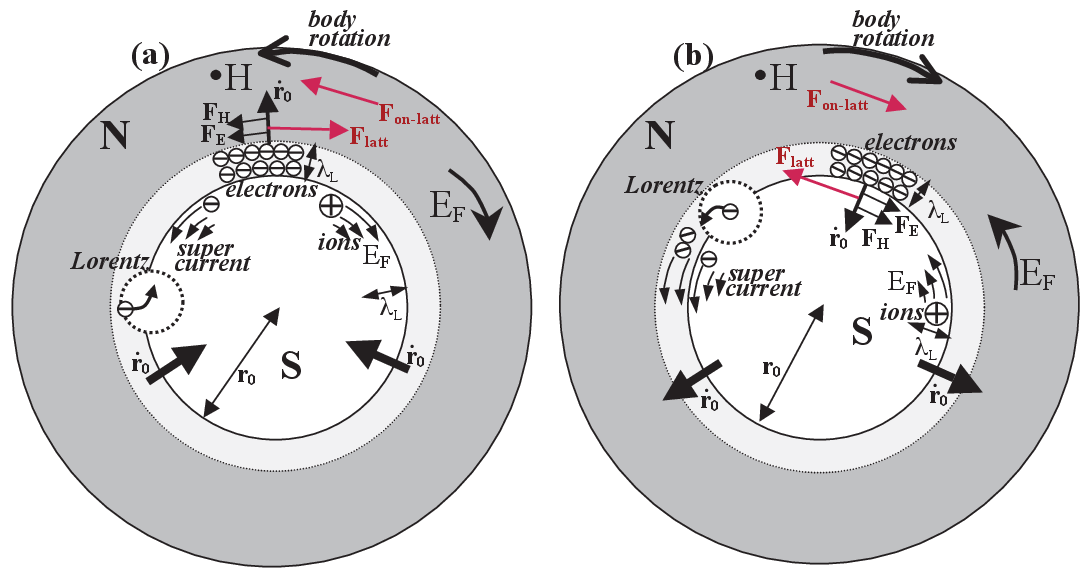}}
 \caption {Figure 14  redrawn replacing the backflowing holes by electrons flowing in the opposite direction. 
 The electric and magnetic forces on electrons
 $F_E$ and $F_H$ point in the same direction. Since the motion is radial, this implies  that another force must exist,
 $F_{latt}$, exerted by the periodic potential of the ions on the charge carriers.
 By Newton's third law, an equal and opposite force is exerted by the charge carriers on the ions,
 $F_{on-latt}$, that makes the body rotate.}
 \label{figure1}
 \end{figure}

In summary, the processes just described qualitatively and analyzed quantitatively in the references \cite{disapp,momentum}
explain how momentum is transferred between electrons and the body as a whole in a reversible way, without any
scattering processes  and   Joule heating that would create entropy in contradiction with theory and experiment.

 \section{closing arguments}
 
 In this paper we have made the simplifying assumption that a sharp boundary between superconducting and normal regions exists,
given by $r_0(t)$, with no supercurrent in the normal region. In other descriptions the superconducting order parameter and the supercurrent
may  decay to zero over a finite distance. However there is no indication for why  in such more complicated descriptions a mechanism
would exist that would allow transfer of momentum from the supercurrent   to the body
without entropy generation and respecting energy conservation, as required by the laws of thermodynamics, within the framework of the conventional theory of superconductivity.

It is important to emphasize that the conclusions reached in this paper about the conventional theory in no way depend on the limitations of BCS theory that relies on weak coupling and long distance averaging. The reason that we are finding a violation of thermodynamics laws within the description given by conventional theory is that conventional theory offers no way to transfer momentum directly from the supercurrent to the body, rather $requires$ an intermediate step where the azimuthal 
momentum is carried by normal electrons \cite{eilen},
$and$ part of the kinetic energy of the supercurrent is carried by normal electrons. Any theory of superconductivity with that requirement will lead to the same violation. The thermodynamic relations teach us that a correct description of superconductivity in nature {\it cannot} involve this intermediate step.

In conclusion, in this paper we have argued that the description of the superconductor-to-normal transition in 
the presence of a magnetic field  and its reverse (Meissner effect) provided by the conventional theory of superconductivity 
\cite{halperin,eilen,scalapino}, which
involves transfer of mechanical momentum between the supercurrent and the body  through an intermediate state involving
normal electrons carrying this momentum, is inconsistent with the laws of thermodynamics.  We conclude that the conventional theory of superconductivity, at least in its present form, 
cannot provide a consistent description of
 the S-N and N-S transitions  in a magnetic field exhibited by all type I superconductors \cite{validity}.
 
 Instead,
the   alternative theory of hole superconductivity   \cite{holesc} 
can, as we briefly discussed in Sect. XVII  and more extensively in references  \cite{revers,momentum,disapp}. 

\acknowledgements
The author is grateful to B. I. Halperin for extensive, detailed and instructive
discussions, to D. J. Scalapino and A. J. Leggett for stimulating discussions, and to
N. Goldenfeld for calling Ref. 25 to his attention.


\begin{references}
\bibitem{referee} This was argued by a referee of an earlier version of this paper.
\bibitem{halperin} B. I. Halperin, private communications. According to Halperin, he does not consider his scenario
      ``to be anything fundamentally new, even if the details have not been spelled out in the literature''. I could not find the details spelled out
      in the literature either.
\end{references}
\end{document}